# Efficient Magnon Injection and Detection via the Orbital Rashba Edelstein Effect

J.A. Mendoza-Rodarte[1,2], M. Cosset-Chéneau[1] *, B.J. van Wees[1] and M.H.D. Guimarães[1]

[1]Zernike Institute for Advanced Materials, University of Groningen, 9747 AG Groningen, The Netherlands
[2]Centro de Nanociencias y Nanotecnología-Universidad Nacional Autónoma de México, Ensenada, 22800-Baja California, México

Orbital currents and accumulation provide a new avenue to boost spintronic effects in nanodevices. Here we use interconversion effects between charge current and orbital angular momentum to demonstrate a dramatic increase in the magnon spin injection and detection efficiencies in nanodevices consisting of a magnetic insulator contacted by Pt/CuO$_x$ electrodes. Moreover, we note distinct variations in efficiency for magnon spin injection and detection, indicating a disparity in the direct and inverse orbital Rashba Edelstein effect efficiencies.

Magnon transport in nonlocal devices [1] is an efficient way of probing the collective excitation of the magnetic order in a variety of materials such as ferrimagnets [2,3], antiferromagnetic oxides [4–6] and van der Waals materials [7]. It permits to access the near equilibrium magnon transport properties in the diffusive regime [8], thus uncovering new type of non-linear phenomenon, such as the magnon diode effect [9,10], while also providing a way to study thermally-driven magnon transport effects [11–13]. Finally, it allows to electrically control a magnon population [14] and opens the possibility to realize magnon transistors [15]. This however relies on the possibility to efficiently inject and detect magnons with a static electrical current using interfacial spin transfer effects [16]. So far, nonlocal magnon transport devices have relied on the direct and inverse spin Hall effect (SHE) [17] in Pt, which has provided only small nonlocal signals, thus limiting the application range of this method. This is particularly problematic for materials with short magnon relaxation lengths, such as van der Waals magnets, since the signals are often too small to explore new functionalities in magnonic devices.

The recent demonstrations [18–24] that spin-charge interconversion effects can be strongly increased by exploiting the orbital angular momentum [25] may however lift this limitation. It has been reported that an orbital current, which is a flux of orbital angular momentum, can be generated with a large efficiency by a charge current through orbital Hall effects in light metals [26,27], or by the direct orbital Rashba Edelstein effects (OREE) at metal oxide interfaces [28–31]. This orbital current is then converted into a spin current by the spin-orbit coupling of either a heavy metal or within the ferromagnet, and is then able to apply a torque on the ferromagnet's magnetization [20,29,32]. The reports that an orbital current can also be efficiently converted into a charge current by the inverse OREE at a Pt/CuOx interface [33,34] open the possibility to achieve large signals in nonlocal magnon transport devices, hence greatly expanding their range of application. This also permits to probe the efficiencies of the direct and inverse OREE in the same device. These efficiencies are defined as the ratio of the generated spin current over the applied charge current for the direct OREE, and as the ratio of the generated charge current over the applied spin current for the inverse OREE. The possibility for these two ratios to be different for interconversion effects taking place at the interface has been heavily debated. [35]. However, nonlocal magnon transport devices exploiting interfacial orbital effects have not yet been explored.

---

*m.n.c.g.cosset-cheneau@rug.nl

Here we demonstrate a dramatic increase in the output signal of yttrium iron garnet ($Y_3Fe_5O_{12}$ – YIG)-based nonlocal magnon transport nanodevices by exploiting the OREE at the Pt/CuO$_x$ interface [29,34,36]. By using the different magnon-generation and detection processes in these devices [Fig. 1(a)], we show that the interconversion by the OREE of a charge current into an orbital current, and of an orbital current into a charge current do not have the same efficiencies.

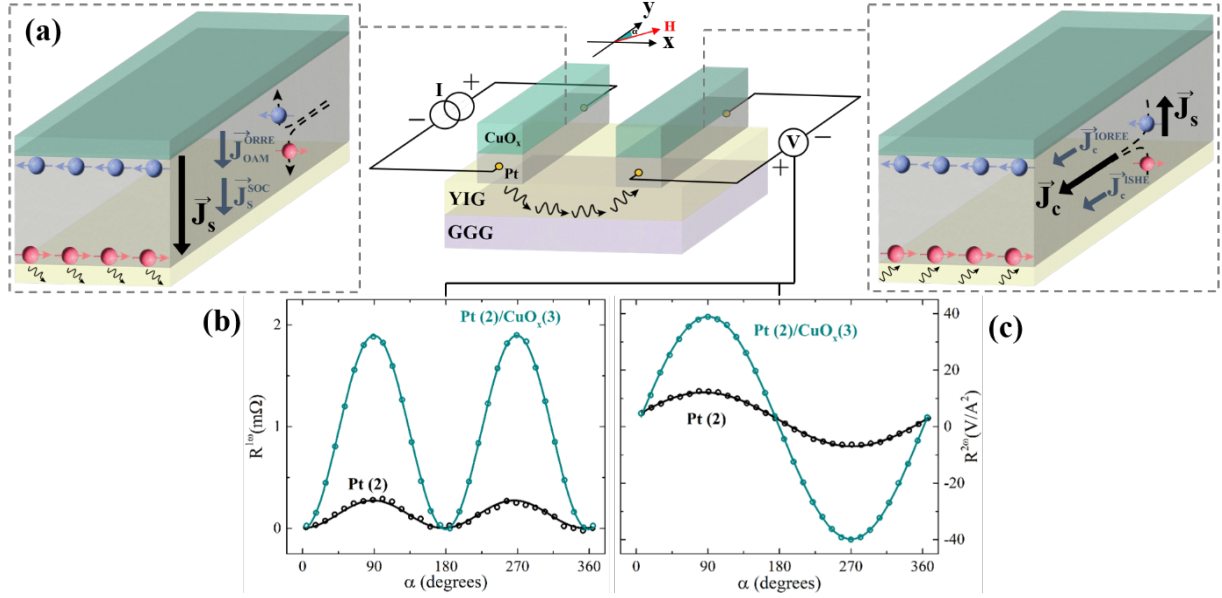

Figure 1: (a) *Left*: Contribution of the direct OREE to the charge to spin interconversion process. The OREE generates an orbital accumulation at the Pt/CuO$_x$ interface, which diffuses as an orbital current ($J_{OAM}^{OREE}$) into the Pt layer. It is then converted into a spin current ($J_s^{SOC}$) and adds up to the one produced by the SHE to form a total spin current $J_s$ which generates a magnon accumulation in YIG. *Center*: Schematic of the nonlocal magnon transport device with its electrical interconnects and the external magnetic field angle $\alpha$. *Right*: Contribution of the inverse OREE to the spin-to-charge interconversion process. The spin current $J_s$ created by the magnon accumulation is converted into a charge current $J_c^{SHE}$ by the SHE in Pt and into an orbital accumulation at the Pt/CuO$_x$ interface. The inverse OREE generates a charge current $J_c^{OREE}$, with both effects resulting in a total charge current $J_c$. First (b) and second (c) harmonic resistance response as a function of $\alpha$. The green and black data points correspond to devices with Pt/CuO$_x$ and Pt electrodes respectively, both with $t_{Pt} = 2$ nm. A baseline has been removed, so that the non-local signals are zero at $\alpha = 0$ and 180 degrees.

The nonlocal magnon transport devices, consisting on two Pt($t_{Pt}$)/CuO$_x$ (3 nm) magnon injecting and detecting electrodes, were fabricated on a 150 nm thick YIG single crystal (Matesy GmbH) grown by liquid phase epitaxy on a $Gd_3Ga_5O_{12}$ (GGG) substrate with (111) orientation. The edge-to-edge distance between the 400 nm-wide 80 μm-long electrodes is fixed for all devices at 2 μm. The devices were fabricated by conventional lithography processes, followed by the d.c. sputtering in Ar$^+$ plasma of Pt and Cu and by a subsequent liftoff. The YIG surface was cleaned using an Ar$^+$ etching step prior to the deposition. The devices were left in ambient conditions for two days to achieve a natural oxidation of the Cu layer. The measurements were performed by rotating an external magnetic field (70 mT) in the plane of the device. We use a conventional lock-in detection technique with a bias current ($I_{bias}$) of 70 μA applied in the injecting electrode at a frequency of 5.1 Hz to measure the first ($V^{1\omega}$) and second ($V^{2\omega}$) nonlocal voltage responses at the detecting electrodes – see Fig. 1(a) for the electrical interconnects. All measurements were performed at room temperature.

Fig. 1(b) and 1(c) show the magnetic field angle ($\alpha$) dependence of the first ($R^{1\omega} = V^{1\omega}/I_{bias}$) and second ($R^{2\omega} = V^{2\omega}/I_{bias}^2$) harmonic nonlocal signals, respectively, measured on devices using Pt(2) and Pt(2)/CuO$_x$(3) electrodes. The signals can be fitted using $R^{1\omega} = R_{nl}^{1\omega}\cos(\alpha)^2$ and $R^{2\omega} = R_{nl}^{2\omega}\cos(\alpha)$, which give the amplitude $R_{nl}^{1\omega}$ and $R_{nl}^{2\omega}$ of the first and second harmonic nonlocal signal [1]. Here we define $R_{nl,Pt}^{1(2)\omega}$ and $R_{nl,CuO_x}^{1(2)\omega}$ as the amplitudes of the first (second) harmonic nonlocal signal measured in devices with Pt only and Pt/CuO$_x$ electrodes, respectively. For $t_{Pt} = 2$ nm, the nonlocal signal amplitudes strongly depend on the presence of the CuO$_x$ layer, with the first harmonic signal going from 0.27 m$\Omega$ for Pt-only electrodes, to 1.9 m$\Omega$ for Pt/CuO$_x$ electrodes – i.e. a factor of 7 increase in the signal. Similarly, the second harmonic nonlocal resistance goes from 9.5 to 39 V·A$^{-2}$ – a factor of 4 increase. When increasing the thickness of Pt, we observe a strong decrease of both $R_{nl,CuO_x}^{1\omega}$ and $R_{nl,CuO_x}^{2\omega}$ [Fig. 2(a) and 2(b)], and the difference between devices with Pt and Pt/CuO$_x$ electrodes becomes less important. This is consistent with the picture discussed in Fig. 1a, where, in order for the OREE to play a role on the magnon injection and detection processes, the thickness of the Pt layer should be much smaller than the spin relaxation length.

To quantify the charge-to-spin interconversion efficiencies as a function of the thickness of the Pt layer, we have to take into account the electrical resistance of the electrodes, since this influences the measured voltage. The nonlocal signals depend on the detecting electrode two point resistance ($R_{2p}$) as $R^{1\omega} = I^{1\omega}R_{2p}/I_{bias}$, and $R^{2\omega} = I^{2\omega}R_{2p}/I_{bias}^2$ with $I^{1(2)\omega}$ being the first (second) harmonic charge current generated by the spin-to-charge interconversion processes at the detecting electrode. In Fig. 2(c) and 2(d) we plot the first and second harmonic nonlocal signals amplitudes normalized by the two point resistance of the detecting electrode, $r_\sigma^{n\omega} = R_{nl,\sigma}^{n\omega}/R_{2p}$ with $n = 1, 2$ and $\sigma =$ Pt or Pt/CuO$_x$. The normalized first and second harmonic nonlocal signals for Pt-only devices, $r_{Pt}^{1\omega}$ and $r_{Pt}^{2\omega}$, display strong drops of nearly an order of magnitude when the Pt thickness is reduced from 3 to 2 nm. This behavior contrasts with the one observed in devices using Pt/CuO$_x$ electrodes, where there is only a slight decrease in $r_{CuO_x}^{1\omega}$ while $r_{CuO_x}^{2\omega}$ increases when $t_{Pt}$ goes from 3 to 2 nm. The difference between the normalized nonlocal signals obtained with and without CuO$_x$ layers vanishes when the Pt thickness increases beyond 3 nm. This directly demonstrates that the presence of a Pt/CuO$_x$ interface leads to an increase of the spin-charge interconversion processes efficiency in the magnon injecting and detecting electrodes with respect to Pt-only electrodes.

The spin-charge interconversion processes in the Pt/CuO$_x$ electrodes possess contributions from the SHE in Pt, and from the charge-orbital interconversion by the OREE at the Pt/CuO$_x$ interface [29,34]. We also point out that even if the 3 nm Cu layer does not fully oxidize in ambient conditions [36], we also expect a strong OREE at the Cu/CuO$_x$ interface [22]. In the injecting electrode, the direct OREE contributes to the charge to spin interconversion by inducing an orbital accumulation at the Pt/CuO$_x$ interface, which then creates an orbital current converted into a spin current by the spin-orbit coupling [37] when flowing into the Pt layer [Fig. 1(a), left panel]. This spin current then crosses the Pt/YIG interface inducing a magnon accumulation under the injecting electrode. After propagation, the magnon accumulation under the detecting electrode creates a spin current which flows into Pt and is partially converted into an orbital current via the high spin-orbit coupling of Pt. The orbital current then creates an orbital accumulation at the Pt/CuO$_x$ interface and subsequently a charge current by the inverse OREE [Fig. 1(a), right panel]. In addition, the SHE in Pt contributes to the spin-to-charge interconversion processes. While the direct and inverse interconversion efficiencies by the SHE were demonstrated to be the same [38], our observations point to the fact that this is not the case for the OREE.

To better quantify the contribution of the direct and inverse OREE to the spin-to-charge interconversion in the Pt/CuO$_x$ electrodes and explore their (different) efficiencies, we make use of the nonlocal signals created by electrically and thermally generated magnons [39]. While the electrically injected magnons depends on the spin-to-charge interconversion efficiency of the injector electrode, thermally injected magnons only depend on the Joule heating generated at the injection point. Since the detection process is the same, the signals from the electrically injected magnons depend on both the injector and detector (i.e. direct and inverse charge-to-spin conversion) efficiencies, while for thermally generated magnons the signals are only dependent on the detection (inverse charge-to-spin conversion) efficiency. These two magnon populations are detected as the first and second harmonic responses [1] with the angular dependences shown in Fig. 1(b) and 1(c), respectively [40].

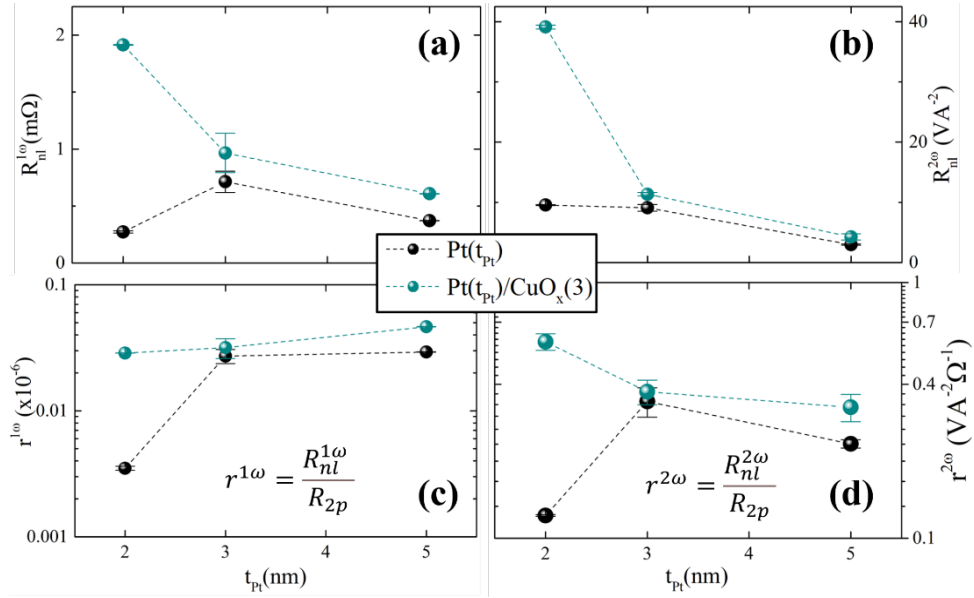

Figure 2: First (a) and second (b) harmonic nonlocal signal amplitudes as a function of the Pt layer thickness ($t_{Pt}$). (c) First and (d) second harmonic nonlocal spin signal amplitudes normalized by the two-point resistance of the detecting electrode as a function of $t_{Pt}$. The green data points correspond to devices with Pt/CuO$_x$ electrodes, while the black data points are for Pt-only electrodes.

We highlight the asymmetry in the injection and detection processes by using a magnon diffusive picture which can be described as follows. The electrical current flowing in the injecting electrode generates a spin current $j_{\text{inj}}^{\text{elec}} = \kappa^D I_{bias}$ which flows across the YIG/Pt interface, with $\kappa^D$ the efficiency of the charge-to-spin interconversion, containing both SHE and OREE contributions. This creates an electrically generated magnon accumulation which diffuses towards the detecting electrode, producing a spin current $j_{\text{det}}^{\text{elec}} = T_{\text{YIG}} j_{\text{inj}}^{\text{elec}}$ at its interface with YIG [8]. For simplicity, here $T_{\text{YIG}}$ contains all the magnon spin transport parameters of YIG and across the YIG/Pt interfaces, such as the magnon conductivity and the spin mixing conductance. The spin current in the detecting electrode ($j_{\text{det}}^{\text{elec}}$), generated by electrical magnon spin injection, is detected as the charge current $I^{1\omega} = \kappa^I j_{\text{det}}^{\text{elec}}$, with $\kappa^I$ being the efficiency of the spin-to-charge interconversion. The thermal gradient caused by the bias current at the injector electrode creates a flow of thermally-generated magnons in YIG, and a spin current $j_{\text{det}}^{\text{therm}} = T_{\text{YIG}}^{\text{therm}} I_{bias}^2$ at the YIG/Pt interface of the detection electrode [11,40]. Here $T_{\text{YIG}}^{\text{therm}}$ is a combination of the thermal transport parameters in YIG and can be used for a linear response of the system to a thermal gradient [41], see the Supplementary Material for details. The thermally generated spin current is converted into a charge current $I^{2\omega} = \kappa^I j_{\text{det}}^{\text{therm}}$

at the detecting electrode. The first and second harmonic normalized nonlocal responses are therefore:

$$r^{1\omega} = \kappa^D T_{YIG} \kappa^I, \quad (1)$$

and

$$r^{2\omega} = T_{YIG}^{therm} \kappa^I, \quad (2)$$

with $\kappa^D$ and $\kappa^I$ the only parameters which depend on the spin-charge interconversion processes in the injecting and detecting electrodes. In the following we note $\kappa_{Pt}^{D(I)}$ the direct (inverse) interconversion efficiencies for the Pt-only electrodes, and $\kappa_{CuO_x}^{D(I)}$ the direct (inverse) interconversion efficiencies for the Pt/CuO$_x$ electrodes. Since the direct and inverse spin-to-charge interconversion effects have the same efficiencies for the SHE, $\kappa_{Pt}^D = \kappa_{Pt}^I$. Finally, the magnon transport properties of YIG, $T_{YIG}$ and $T_{YIG}^{therm}$, are assumed to be independent of the presence of a CuO$_x$ layer on the electrodes.

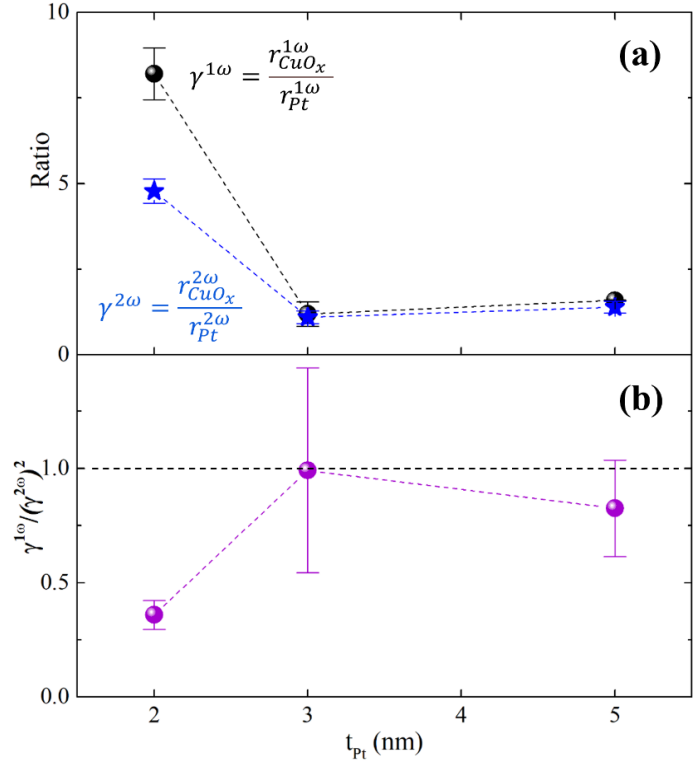

Figure 3: (a) $\gamma^{1\omega}$ (black circles) and $\gamma^{2\omega}$ (blue stars) and (b) $\gamma^{1\omega}/(\gamma^{2\omega})^2$ as function of $t_{Pt}$.

The spin-to-charge interconversion efficiencies in the Pt and Pt/CuO$_x$ electrodes for injection and detection can be directly compared by plotting $\gamma^{1\omega} = r_{CuO_x}^{1\omega}/r_{Pt}^{1\omega}$ and $\gamma^{2\omega} = r_{CuO_x}^{2\omega}/r_{Pt}^{2\omega}$ [Fig. 3(a)]. We observer that the charge current generated at the detecting electrode by the electrically generated magnons ($\gamma^{1\omega}$) shows an increase of a factor of 8 for $t_{Pt} = 2$ nm in the presence of a CuO$_x$ capping layer with respect to uncapped electrodes. In contrast, the charge current at the detector electrode obtained by thermally generated magnons ($\gamma^{2\omega}$) show an increase of a factor of 5. No significant increase of the generated first and second harmonic charge current is observed for $t_{Pt} = 3$ and 5 nm, and $\gamma^{1\omega} \approx \gamma^{2\omega} \sim 1$. This indicates that for a Pt thickness larger than its spin relaxation length ($\sim 2$ nm), the presence of a CuO$_x$ capping layer has no effect on our signals, indicating that the OREE does not contribute to the spin-to-charge interconversion. This is in agreement with previous reports in other types of measurements [29,34,36]. Nonetheless, none of these reports have been able to detect and separate the direct and inverse OREE in the same device.

The electrical magnon spin injection depends on both the direct and inverse charge-to-spin interconversion efficiencies, while the signals generated by thermally injected magnons depends only on the inverse interconversion efficiency. The efficiencies of the direct and inverse charge-to-spin interconversion processes can be quantified by the ratio:

$$\frac{\gamma^{1\omega}}{(\gamma^{2\omega})^2} = \frac{\kappa^D_{CuO_x}}{\kappa^I_{CuO_x}}.$$

If we assume a similar efficiency for the direct ($\kappa^D_{CuO_x}$) and inverse ($\kappa^I_{CuO_x}$) OREE-assisted spin-charge interconversion effects, this implies that $\gamma^{1\omega} = (\gamma^{2\omega})^2$ or $\kappa^D_{CuO_x} = \kappa^I_{CuO_x}$. This is in stark contrast with what we observe for our devices with $t_{Pt} = 2$ nm [Fig. 3(b)] where $\gamma^{1\omega}/(\gamma^{2\omega})^2 = 0.4$, clearly indicating an asymmetry in the injection and detection processes in these devices.

We thus conclude that the efficiency of the orbital-to-charge interconversion by the OREE is larger than the efficiency of the charge-to-orbital interconversion by this same effect, i.e. $\kappa^I_{CuO_x} = 0.4\, \kappa^D_{CuO_x}$. This situation is reminiscent of the previously observed different efficiencies of the spin-charge interconversions by the Rashba Edelstein effect at a metal oxide interface [35], which was attributed to the different scattering times within the interfacial states and in the metallic bulk. Our results indicate that a similar mechanism might be at play in the OREE at metal/oxide interfaces.

We envision that the use of the OREE to increase nonlocal magnon signals by nearly one order of magnitude, as we show here, can be used to expand the range of materials for magnon transport. This includes, for example, materials with short magnon relaxation lengths for which nonlocal magnon signals tend to be very small, but might still show interesting magnonic effects. Furthermore, the possibility of generating very large magnon accumulations should give easier access to nonlinear magnonic phenomena such as magnon condensates [42] and spin diodes [9]. Finally, the use of the orbital degree-of-freedom can also be used to study the transport of collective orbital excitations in insulating materials [43] using nonlocal devices.

**Acknowledgement**

We acknowledge the technical support of J. Holstein, H. de Vries, F.H. van der Velde, H. Adema, and A. Joshua. This work was supported by the Dutch Research Council (NWO – OCENW.XL21.XL21.058), the Zernike Institute for Advanced Materials, the research program "Materials for the Quantum Age" (QuMat, registration number 024.005.006), which is part of the Gravitation program financed by the Dutch Ministry of Education, Culture and Science (OCW), and the European Union (ERC, 2D-OPTOSPIN, 101076932, and 2DMAGSPIN, 101053054). Views and opinions expressed are however those of the author(s) only and do not necessarily reflect those of the European Union or the European Research Council. Neither the European Union nor the granting authority can be held responsible for them. J.A.M.R. acknowledges the financial support of CONACYT. The device fabrication and characterization were performed using Zernike NanoLabNL facilities.